\documentclass[nofootinbib,twocolumn,showpacs,amssymb,floatfix,aps]{revtex4-2}
\pdfoutput=1

\usepackage [latin1]{inputenc}
\usepackage{graphicx}
\usepackage{dcolumn}
\usepackage[english]{babel}
\usepackage{amsmath}

\begin{document}

\preprint{}
\title{
High-energy photons from Gamma-Ray Bursts, but no neutrinos} 

\author{A. De R\'ujula${}^{a,b}$}
\affiliation{  \vspace{3mm}
${}^a$Instituto de F\'isica Te\'orica (UAM/CSIC), Univ. Aut\'onoma de Madrid, Spain;\\
${}^c$Theory Division, CERN, CH 1211 Geneva 23, Switzerland
}

\date{\today}
             
 \begin{abstract}
 
 The Cannon-Ball model of Gamma-Ray Bursts and their afterglows --described
 in the text and in innumerable previous occasions-- is extremely successful and
 predictive. In a few intrinsically bright GRBs, gamma-rays with energies in the TeV range
have been observed. The CB model, I argue, has no difficulty in describing the origin and
approximate properties of these high-energy gamma rays and the extreme difficulty
of observing their accompanying neutrinos.

\end{abstract}


\maketitle

\section{Introduction}

In a small fraction of intrinsically bright (Long) Gamma-Ray Bursts (GRBs) gamma-rays 
with energies in the TeV range
have been observed: GRB180720B by H.E.S.S. \cite{Abdalla-etal:2019}, 
190114C by MAGIC \cite{Acciari-etal:2019}, 
190829A by H.E.S.S. \cite{Abdalla-etal:2021}, 
201216C by MAGIC \cite{Abe-etal:2024} 221009A (the brightest so far) 
by LHAASO \cite{Cao-etal}. 

How are the above mentioned very high-energy (VHE) $\gamma$'s produced? 
Since VHE $\gamma$'s and neutrinos are expected to be co-generated by similar mechanisms
\cite{OldNeutrino-expectations}, why have the GRB-associated neutrinos not been seen?
The ``CannonBall" (CB) model of GRBs and Cosmic Rays (CRs) \cite{CBoldies}
provides very simple answers to the above questions.

\section{A few relevant observations}

\begin{figure}[]
\centering
\includegraphics[width=8.6cm]{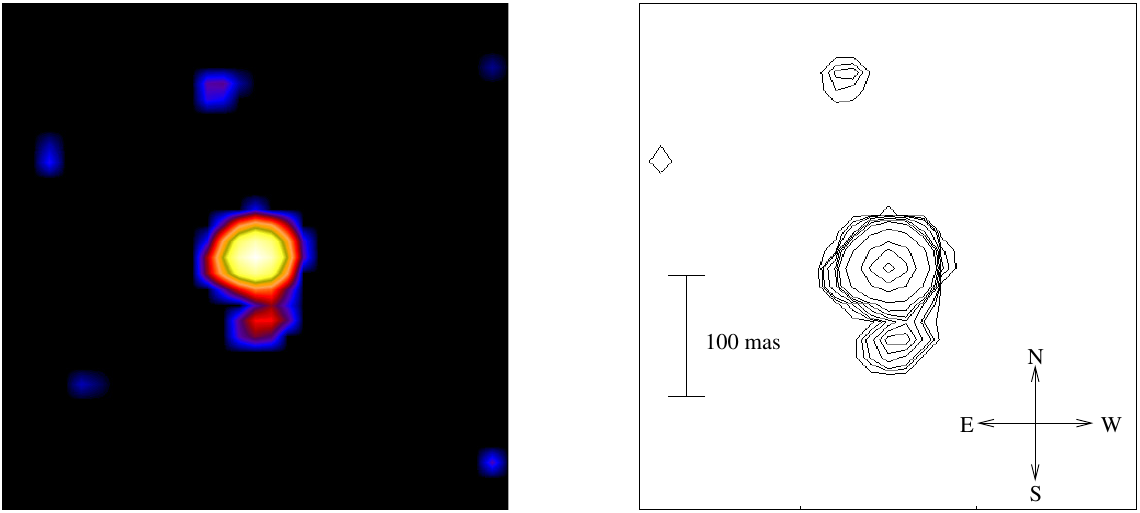}
\caption{New reconstructed image (left) and contour plot (right) of a 300
mas field around SN1987A from speckle data recorded with a 10 nm Wide
Filter centered at 653.6 nm. Note that two of the noise spikes (in the lower
left and right) do not show up in the contour plot because their intensity
falls below the lowest contour level (1\%).}
\vspace{-1cm}
\label{Fig:Papaliolios}
\end{figure}

Some core-collapse supernovae (SNe) emit highly relativistic ``cannonballs". A first and remarkable such observation was that of SN1987A. As stated by Ninelson and Papaliolios in the abstract of their second paper on the subject: {\it If the spots were ejected from the SN then the velocities of the spots are relativistic and the 2nd spot appears to be superluminal and must be blue-shifted} 
\cite{NinAndPap}. The figure in their paper is so historically relevant that I cannot resist reproducing it with its caption: Fig.(\ref{Fig:Papaliolios}).

Another relevant case is that of GRB030329 for which Taylor et al.$\!$ \cite{Taylor}
remark in their abstract: {\it Much less easy to explain is the single observation 52 days after the burst of an additional radio component 0.28 mas northeast of the main AG. This component requires a high average velocity of 19c and cannot be readily explained by any of the standard models.} Surprisingly, they did not publish a figure of this groundbreaking result.

The merger of two neutron stars (NSs) 
\cite{EarlyNSNSmergers}
also results in the emission of CBs \cite{ShavivDar,ADRinNC}. In the historical case of
the Gravitational Wave and Short Hard (gamma ray) Burst GW+SHB 170817 a CB was seen to move at a average superluminal velocity of $(4.0\pm 0.4)$c between days 75 and 230 after burst \cite{Mooley}.
 In this and the previously cited cases the data were insufficient to resolve the CBs, they are effectively point-like\footnote{``Seeing is believing" is not always applicable. In the case of the SHB170817, for instance, what the observers first saw was {\it A successful jet that drives a cocoon through interaction with the dynamical ejecta.} The interpretation changed in their successive articles.
 Jets have {\it not} been {\it seen} in GRBs nor in NS-NS mergers.}.

\section{The CB model }

A satisfactory description of how core-collapse SNe implode and explode does not yet exist. 
How these stars and the NS-NS mergers resulting in SHBs emit CBs is also not understood. But since cannonballs are observed, one might as well ask about their effects. They provide the basis for a
very simple and quite predictive understanding of GRBs, X-Ray Flashes (XRFs) and CRs. This fact 
is not recognized by the defenders of the respective 
``standard" models\footnote{There are realms of physics, such as cosmology and high-energy physics, whose standard models are predictive and successful. The name of the game is to challenge them, not the contrary.}, but is acknowledged by AI \cite{AI}, which (who?) appreciates the 
CB model's predictive power.
How the CB model, the  ``standard" theories of GRBs and the data confront one another is discussed in detail in
\cite{DDDcompare} and summarized in its tables II and III. Conventional ideas on CRs are
compared with data in \cite{CRpdg}. 

The CannonBall model is based on the assumption that Gamma Ray Bursts, X-ray Flashes (XRFs) and Cosmic Rays are generated by the CBs that striped-envelope core-collapse supernovae 
(SNe) emit \cite{DD2008}. The CBs are made of ordinary matter and generate GRBs --one pulse per approaching CB-- as their electrons inverse-Compton-scatter (ICS) the light of their parent SN's {\it glory}: the {\it wind} of pre-SN
ejections, temporarily illuminated during the explosion by the SN's early UV light.

The nuclei and electrons of the Inter-Stellar Medium (ISM) are scattered by the CBs' magnetic field becoming the primary non-solar Galactic CRs, including nuclei and electrons. Below their corresponding CR ``knees'' the scattering is elastic, the extremely tiny fractions of CRs above the individual knees have been Fermi-accelerated within the CBs. For recent updates, see \cite{DDupdate,ADRupdate}.

\subsection{GRB peak energies}

\begin{figure}[]
\centering
\includegraphics[width=8.6cm,height=6.6cm]{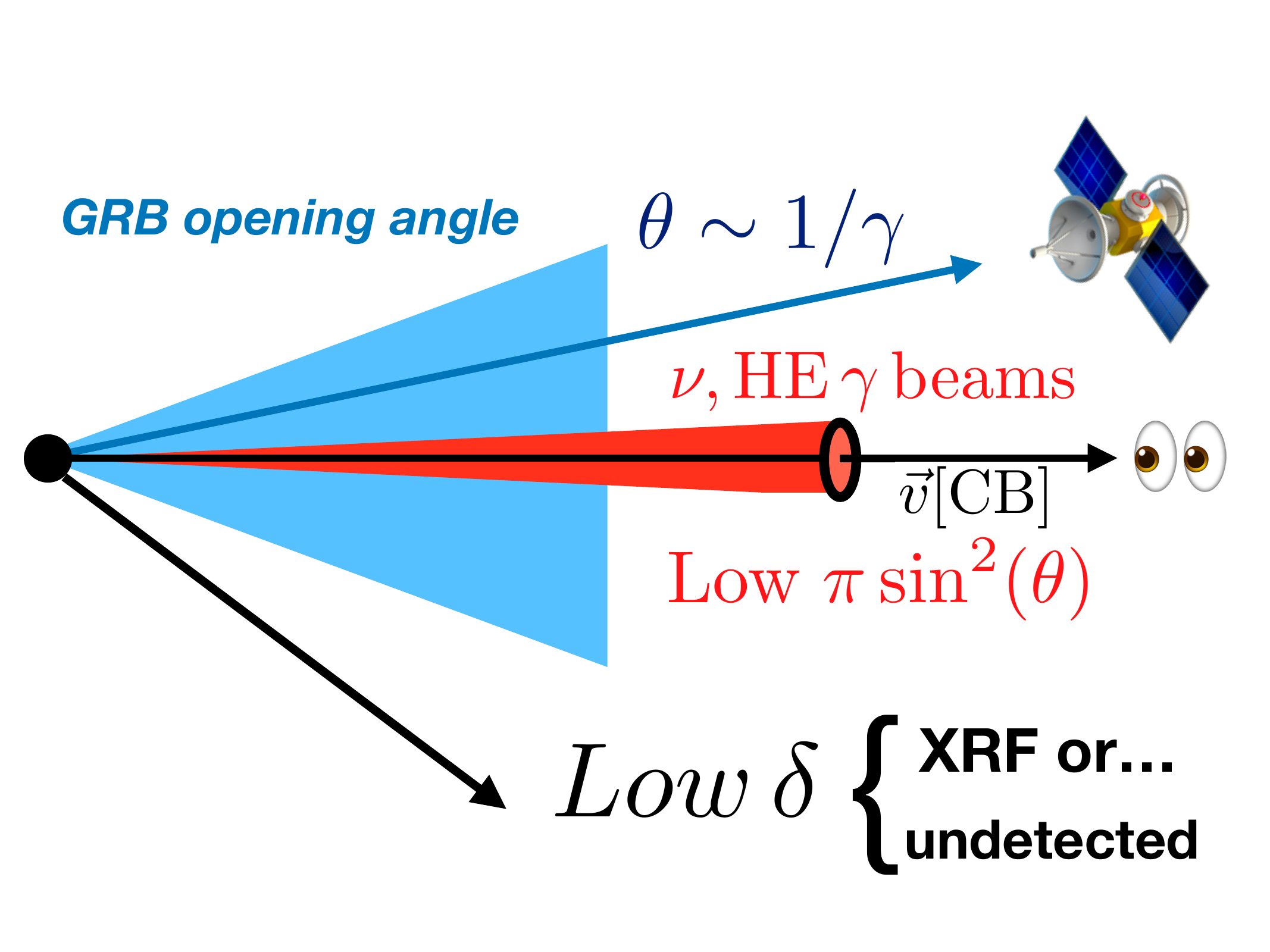}
\vspace{-1.2cm}
\caption{Opening angles of a conventional GRB (the blue range) and of an XRF
(the black arrow),
compared to the ones of the high-energy beams of $\gamma$ rays and neutrinos.
An observer is very unlikely to be where the eyes are. }
\vspace{-.6cm}
\label{Fig:Angles}
\end{figure}

Let $\gamma$ be the Lorentz factor of a CB moving at an angle $\theta$ relative to the line of sight, as in Fig.(\ref{Fig:Angles}). The Doppler factor by which the CB's IC-scattered radiation is (locally) boosted is:
\begin{equation}
           \delta = {1/ [\gamma\, (1-\beta\, \cos\theta)]}\approx 
           {2\gamma/ (1+\gamma^2\, \theta^2)},     
\label{Doppler}  
\end{equation}
where the approximation is valid for $\gamma^2 \gg 1$ and $\theta^2 \ll 1$, 
the domain of interest here. 

\begin{figure}[]
\centering
\vspace{-3cm}
\includegraphics[width=9cm]{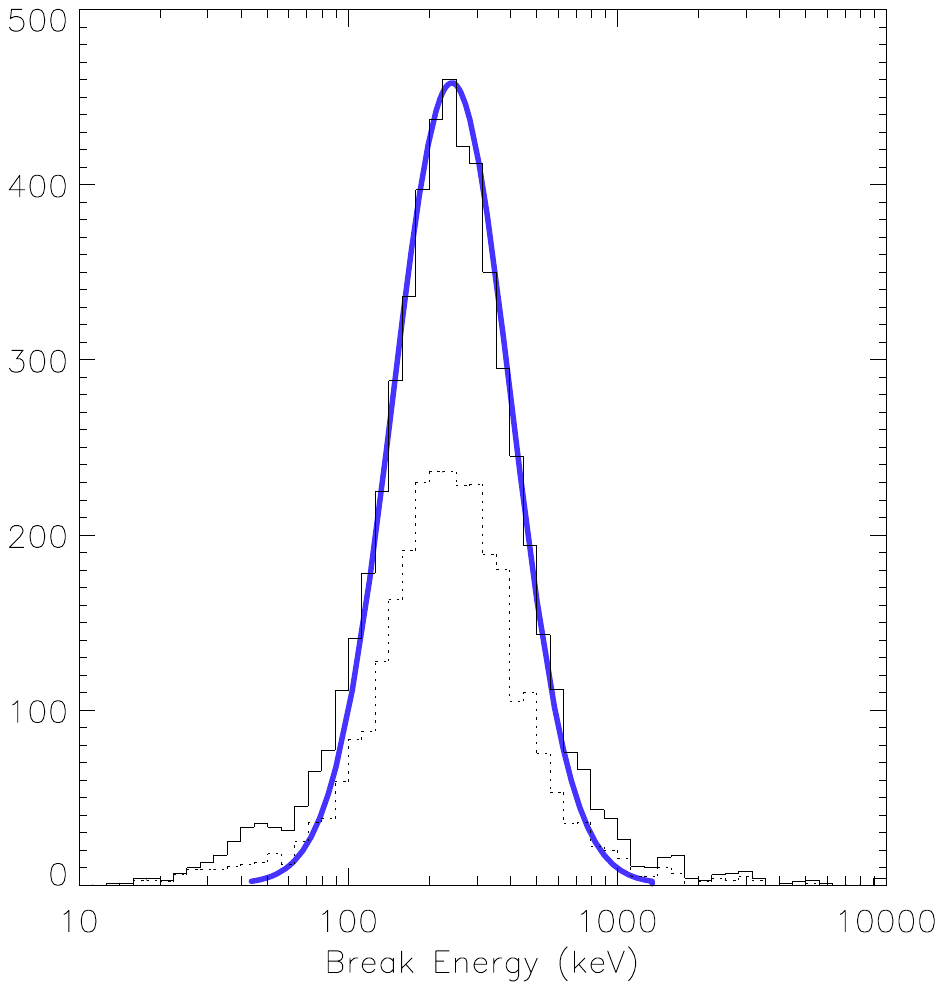}
\vspace{-1.5cm}
\caption{Peak or ``break" energy distribution of long GRBs \cite{Preece}. 
The blue curve is independently
extracted from the CB-model analysis of GRB afterglows
\cite{DD2004}.}
\vspace{-2cm}
\label{Fig:Epobs}
\end{figure}

Consider an electron, comoving with a CB traveling with a Lorentz factor $\gamma$, and
a glory's photon of energy $E_i$ moving at an angle $\theta_i$ relative to the CB's direction of motion.
They Compton-scatter. 
The outgoing photon energy is totally determined:
\begin{equation} 
\begin{gathered}
E_p \!=\!  {\gamma\,\delta\over 1\!+\!z}\,
(1\!+\!\cos\theta_i) E_i 
\!=\!  (250\;{\rm keV})\,
\sigma {1\!+\!\cos\theta_i\over 1/2}
{E_i\over 1\;\rm eV}\\
\sigma\! \equiv\! {\gamma\,\delta\over 10^6}\,{2\over 1\!+\!z},
\end{gathered}
\label{Eq:boosting} 
\end{equation}
where I set $\beta\!\approx\! 1$ and, for a {\it semi}~transparent wind, 
$\langle\cos\theta_i\rangle\!\sim\!-1/2$. The result in Eq.(\ref{Eq:boosting}) is normalized
to a typical $\gamma\!=\!10^{-3}$, $\theta\!=\!1/\gamma$ and, 
for pre-Swift GRBs, $\langle z \rangle\!\approx\! 1$. The glory's light may
be roughly approximated by a thin-thermal
spectrum \cite{Falk} with a peak energy $\sim\! 1$ eV. Thus 
$E_p\!\simeq\!250$ keV is the average peak or `break' energy 
in Eq.~(\ref{Eq:boosting}) and Fig.(\ref{Fig:Epobs}).

The above explanation is not a prediction, but the distribution of peak energies {\it is}
a CB-model prediction. The histogramed $E_p$
results shown in Fig.(\ref{Fig:Epobs}) are pre-Swift measurements \cite{Preece}.
The continuous curve displays in blue, with use of Eq.(\ref{Eq:boosting}) and $\theta\sim 1/\gamma$,
 a fit to measured initial $\gamma$ values extracted from fits \cite{DD2004}
to GRB {\it afterglows} (AGs). These are generated by synchrotron radiation and do not involve $E_p$
measurements.

Many GRB observations are even simpler to explain in the CB model. The peak energy $E_p$ and the ``isotropic equivalent energy'' $E_{iso}$ are, respectively, proportional to $\gamma\delta/(1+z)$ and 
$\gamma\delta^3$.
For typical viewing angles $\delta\!\approx\!\gamma$, hence $(1+z)E_p\!\propto\! E_{iso}^{1/2}$; this is the ``Amati'' correlation \cite{Amati}. The peak luminosity is $(1+z)^2 L_p\!\propto\! \delta^4$. Thus $E_{iso}\!=\![(1+z) L_p]^{3/4}$. This is the ``Yonekotu'' correlation \cite{Yonetoku}. 
Both correlations are satisfied by the
data, as well as about a dozen other CB-model correlations \cite{DDDcorrels}.

In the CB model the energy distribution in a GRB pulse is also understood. Describing again 
the early glory's light energy
distribution \cite{DD2004} as a thin-thermal spectrum  $dN/dE\propto {\rm Exp}[-E_g/T_g]/E_g$, with $T_g$ the pseudo-temperature. Boosted by ICS the pulse's energy distribution and spectral energy density
become\footnote{There is a more refined and successful CB-model account for $dN/(dE\,dt)$ including
the time-dependence of the energy distribution in a CB pulse \cite{DDDcompare}. In standard models the shape of $dN/dE$ is parametrized by a ``Band" function \cite{Band}, an arbitrary choice with three unmeaning parameters to be fit.}:
\begin{equation}
\begin{gathered}
dN/dE\propto {\rm Exp}[-E/T]/E,\;\;\; T={\gamma\,\delta \over (1+z)}\, T_g\\
\nu \, dN/d\nu=E\,dN/dE\vert_{E\to h\,\nu}.
\end{gathered}
\label{Eq:Edistrib}  
\end{equation}

\section{A hint}

The ``classic'' results of Fig.(\ref{Fig:Epobs}) characterize most long-duration GRBs, but there
are cases of very luminous GRBs with $E_p$ significantly higher than the mean. Some of the ones listed in \cite{Ravasio}, have $E_p$ values estimated to be 
$\sim \! 2400,\,10560,\,5320$ keV 
(GRBs 110721, 110328, 151006, respectively). For a typical $\gamma\!=\!10^3$, even for 
$\theta\!\ll\! 1/\gamma$ --so that $\delta\!\simeq\! 2 \gamma$ in Eqs.(\ref{Eq:boosting})-- these values are high. Observed GRBs with very large $E_p$ are infrequent (the opening angle of their $\gamma$-ray beam is significantly smaller than average) and they have Lorentz factors significantly higher than a typical $10^3$. None of this is extraordinary, but it is a hint of what to expect for VHE GRBs.

In the prompt radiation of the VHE GRB221009A \cite{Axelsson} successive time intervals of 
$\cal O$(1)-second duration have $E_p$ values  $\sim\! 15, \, 7, \, 6$ MeV. This is what is expected for successive CBs with very very high $\gamma$ ($\gg 10^3$), observable only at very very small $\theta$, all as in Fig.(\ref{Fig:Angles}). 

Understanding the $E_\gamma={\cal O}(1)$ TeV VHE $\gamma$ rays --which are made in the 
CB model (as of this article) 
by collisions between nuclei in the CB and ambient nuclei-- requires input from laboratory data, discussed in subsection \ref{ss:LHCdata}.

\section{VHE positrons and photons}
\label{s:VHE}

\subsection{The ``wind"}
\label{ss:wind}

Massive stars lose mass in the form of ``winds", before they die in SN explosions. 
We  refer to the pre-SN close-by material, accumulated by previous 
ejecta, as ``the wind". As discussed in great detail in \cite{DD2008}, 
the observations 
\cite{winds, winds2} 
indicate very high wind particle-number densities, 
$n\!\sim\! 5 \!\times\! 10^{7}$ cm$^{-3}$, at the distances, $l\!=\! {\cal{O}}(10^{16})\,\rm cm$
 of interest to the production of GRBs and positrons in the CB model. The measured 
 number density, $n$,
and mass density,  $\rho$,
decline roughly as $1/l^2$ and the wind's  {\it ``surface density''} is 
$\Sigma\!\equiv\! \rho\, l^2\! \sim\! 10^{16}\; {\rm g\, cm^{-1}}$ \cite{winds,winds2}.

The CB-model input priors are shown in Table I.

\begin{table}
      \caption[]{\bf Input priors of a CB and the SN's
      ``wind''.}
             \begin{tabular}{cll}
            \hline
            \noalign{\smallskip}
            Parameter     &  \rm Value & Definition \\
            \noalign{\smallskip}
            \hline
            \noalign{\smallskip}
 $\gamma$      & ${\cal O}(10^3)$ & CB's Lorentz factor$^{\mathrm{a,b}}$ \\
 $N_{\!B}$ & $10^{50}$ & CB's baryon  number$^{\mathrm{a}}$ \\
 $c_s$ & $c/\sqrt{3}$& CB's expansion velocity$^{\mathrm{a}}$ \\
 $\Sigma$ & $10^{16} {\rm g/cm}$& Wind's surface density$^{\mathrm{a}}$ \\
\hline
         \end{tabular}
\begin{list}{}{}
\item[$^{\mathrm{a}}$] Typical CB-model value \cite{DD2004}.
\vspace{-.15cm}
\item[$^{\mathrm{b}}$] The $\gamma$ distribution is that of \cite{DD2004}, here Eq.(\ref{eq:Dgamma}).
\end{list}
   \end{table}

\subsection{A CB sailing in the wind}

In what follows, to avoid pedantic factors close to unity, we consider the
composition of a CB (but not of the SN's wind) to be that of hydrogen. For protons with 
$\gamma\!=\!{\cal O}(10^3 \, {\rm to}\,10^4)$,
the $pp$ total cross section on a fixed target, $\sigma_{pp}\!\approx\! 40$ mb, is dominantly inelastic.
The $p$-nucleus cross section on light nuclei is, to a sufficient approximation,
an incoherent sum over the $p$-nucleon cross sections, with $\sigma_{pn\!}\approx\!\sigma_{pp}$.
Consequently, for a light nucleus ${\cal \small  N}(A)$ of mass number $A$,
$\sigma_{p \cal \small  N}\!\sim\! A\,\sigma_{pp}$.
Thus, for a wind's material dominated by the light elements of the pre-supernova surface
we do not need to know its precise composition.


An expanding CB becomes transparent to the ambient nucleons it hits when their attenuation length becomes of the order of the CB's radius. Thus, nuclear collisional transparency occurs at
$R_{pp}\!\sim \![3 \,\sigma_{pp}\,N_{\!B}/(4\pi)]^{1/2}\!\sim\! 10^{12}$ cm,
for  $N_{\! B}\!=\! 10^{50}$. By assumption a CB initially expands (in its rest frame) at a radial velocity 
 $c_s\!=\!c/\sqrt{3}$: the speed of sound in a relativistic plasma.
When the $pp$ collisions cease the CB has travelled a distance 
$l_{max}(\gamma)\!=\!\sqrt{3}\,\gamma\,R_{pp}\!=\!1.7\times10^{15}\,(\gamma/10^3)\,{\rm cm}$,
where  $\gamma$ relates  the times in the CB and SN rest systems.

What fraction of the CB's protons is lost to interactions with the wind?
Let the transverse radius of a CB at a distance $l$ from the SN be $r_{_{\!\rm CB}}(l)$, corresponding
to a surface $S(l)=\pi\,r_{_{\!\rm CB}}^2=\pi\,l^2/(3\,\gamma^2)$. The wind's baryon-number density 
is $n(l)\approx \Sigma/(m_p\,l^2)$, with $m_p$ the proton's  mass. The number of 
$pp$ plus $pn$ collisions ($p{\cal \small  N}$) is:
\begin{equation}
N_{p{\cal \small  N}}\!=\!\int_0^{l_{max}(\gamma)} S\,n\, dl\!=\!{\pi\,R_{pp}\over{\sqrt{3}\,\gamma}}\,{\Sigma\over m_p}
\!=\!1.1\times 10^{49} \,{10^3\over\gamma},
\label{Npp}
\end{equation}
so that a CB with $\gamma=10^3$ would have lost
$\sim 10\%$ of its $\sim 10^{50}$ protons to $p$-wind interactions, a negligible effect.

\subsection{Surviving attenuation by the wind}
\label{ss:attenuation}

 For the rest of section \ref{s:VHE} the discussion is common to the VHE GRB gamma rays and 
 the CR positron flux excess \cite{AMSPos} observed above the expectation from 
 secondary positron production. Thus I repeat to some extent the derivation in
\cite{ADReplus}, that dealt exclusively with positrons.

The $p$-wind collisions give rise, mainly via the chain 
$pp\,({\rm or} \, pn)\!\to\! \pi\,{\rm or}\, K\!\to\! \mu^+\!\to\! e^+$, to a CR primary source flux of positrons
\cite{ADReplus}. Similarly, 
the chains $pp\,({\rm or} \, pn)\!\to\! \pi^0\!\to\!\gamma\gamma$ and  $pp\,({\rm or} \, pn)\!\to\!\eta$ and  $\eta\!\to\!\gamma s$ are the dominant sources of VHE photons, except in the extremely forward direction, as discussed in detail in Section \ref{Sec:Nonus}.

Not all  photons and positrons manage to penetrate the SN's wind environment.
Let $\sigma_T=0.665\times 10^{-24}$ cm$^{2}$ be the Thomson cross section,
 an adequate constant numerical value approximating very well the $\gamma$ (or $e^+$) attenuation by the wind material,  dominantly due to Thomson scattering and $e^+\,e^-$ pair creation on nuclei and electrons, see figures 34.15 and 34.16 in \cite{pdg}.

Not all the photons and positrons made in $p$-wind collisions escape
unimpaired to become observable: some are absorbed by the wind.
The probability that a $\gamma$ or an $e^+$ produced at a distance $l$ from the SN
evades this fate, in a wind with a density profile $n_e\propto l^{-2}$,
is $A(l)= {\rm exp}[-(l^w_{tr}/l)^2]$ with
$l^w_{tr}=\sigma_T\, \Sigma/m_p$  the $l$ at which the 
remaining ``optical" depth of the wind is unity. 

Still referring to a single CB with $N_{\! B}=10^{50}$ and LF $\gamma$, let us
estimate the number, $N_{\rm out}$, of proton-wind collisions whose produced $\gamma$-rays
and positrons
penetrate the wind unscathed. To do so, add an extra factor $A(l)$ to the integrand in 
Eq.(\ref{Npp}) and integrate. The result is:
\begin{eqnarray}
&&N_{\rm out}(\gamma)={\pi\,R_{pp}\over{3\,\gamma^2}}\,{\Sigma\over m_p}\,I[l^w_{tr},l_{max}(\gamma)],
\nonumber\\
\!\!\!\!I&\equiv& l_{max}\,{\rm Exp}[-(l^w_{tr}/l_{max})^2]-\sqrt{\pi}\,l^w_{tr}\,{\rm erfc}[l^w_{tr}/l_{max}].
\label{Nppe}
\end{eqnarray}

\subsection{The Lorenz factor distribution of CBs}
\label{ss:CBgamma}

Next, we ought to weigh the above fixed LF
result with the distribution, $D(\gamma)$, of the Lorentz factors of 
CBs, the one depicted in Fig.(\ref{Fig:Epobs}), extracted from GRB afterglows \cite{DD2004}
and used to predict the CR knee energies of H, He, Fe and electrons \cite{DD2008, DDupdate}:
\begin{eqnarray}
&&D[\gamma]={\rm exp}(-[(y-y_0)/\sigma]^2);\nonumber\\
&&{\rm where}\,y\equiv {\rm Log}_{10}[\gamma^2],\;\;y_0=6.3,\;\;\sigma=0.5;
\label{eq:Dgamma}
\end{eqnarray}
\vspace{-.5cm}
so that
\begin{equation}
\bar N_{\rm out}(\gamma)=D[\gamma] \,N_{\rm out}(\gamma).
\label{eq:Dbar}
\end{equation}

\begin{figure}[]
\vspace{-0.4cm}
\hspace{-.5cm}
\centering
\includegraphics[width=9cm]{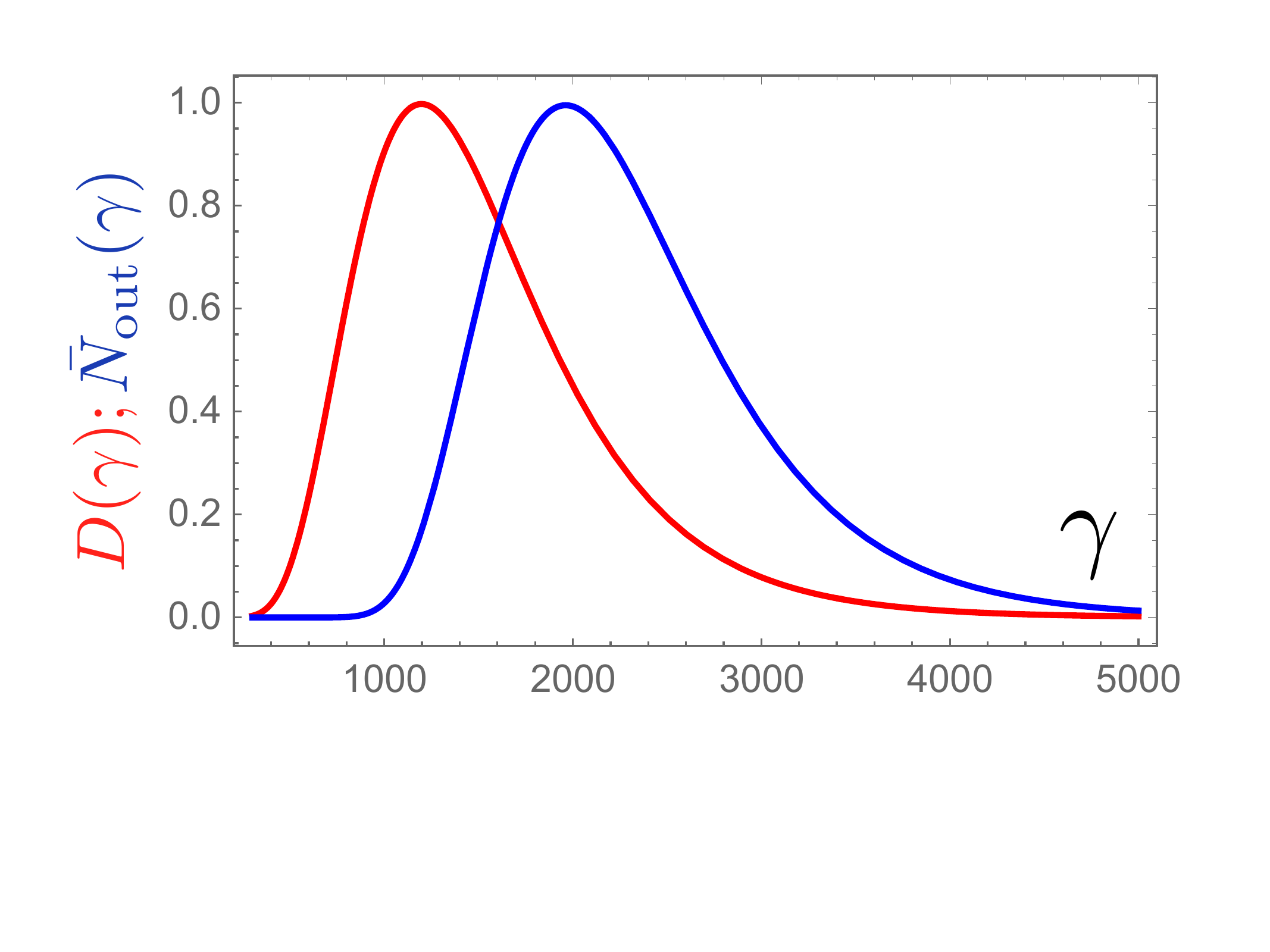}
\vspace{-2.3cm}
\caption{In red $D(\gamma)$, Eq.(\ref{eq:Dgamma}). In blue $\bar N_{\rm out}(\gamma)$,
Eq.(\ref{eq:Dbar}). Both distribution are arbitrarily
normalized in the figure.}
\vspace{-0.4cm}
\label{fig:GammaDistrs}
\end{figure}

To further clarify: the function 
$\bar N_{\rm out}(\gamma)$ is the number distribution of proton-wind collisions 
resulting in $\gamma$ rays that penetrate the wind, for a single CB of $N_B=10^{50}$, and $\gamma$
randomly picked from the distribution $D(\gamma)$. The shapes of $D(\gamma)$ and
$\bar N_{\rm out}(\gamma)$ are drawn in Fig.(\ref{fig:GammaDistrs}), showing how
$\bar N_{\rm out}(\gamma)$ is weighed to higher LFs than $D(\gamma)$ because larger-$\gamma$ CBs 
keep interacting with the wind up to distances at which the latter is getting thinner.

The distribution $D(\gamma)$ peaks at $\gamma\!\simeq\! 1565$ and its mean is 
$\bar \gamma\!\simeq\! 1811$, while $N_{\rm out}(\gamma)$ peaks 
at $\gamma\!\simeq\! 1950$ and its mean is 
$\gamma_{\rm mean}\!=\!2496$. CB's protons --which made photons exiting the wind with
 a comoving LF $\gamma_{\rm mean}$--
have a mean energy $E_p[CB]\!=\!2.34$ TeV. Excellent news, since that is the energy domain at
which these protons may make the VHE $\gamma$ rays observed in some GRBs.
To discuss the energy and angular distributions of these $\gamma$ rays, inputs 
from laboratory experiments are needed. Before discussing them, a brief reminder:

\subsubsection{A Positron interlude}
\begin{figure}[]
\vspace{-.1cm}
\hspace{-.5cm}
\centering
\includegraphics[width=9cm]{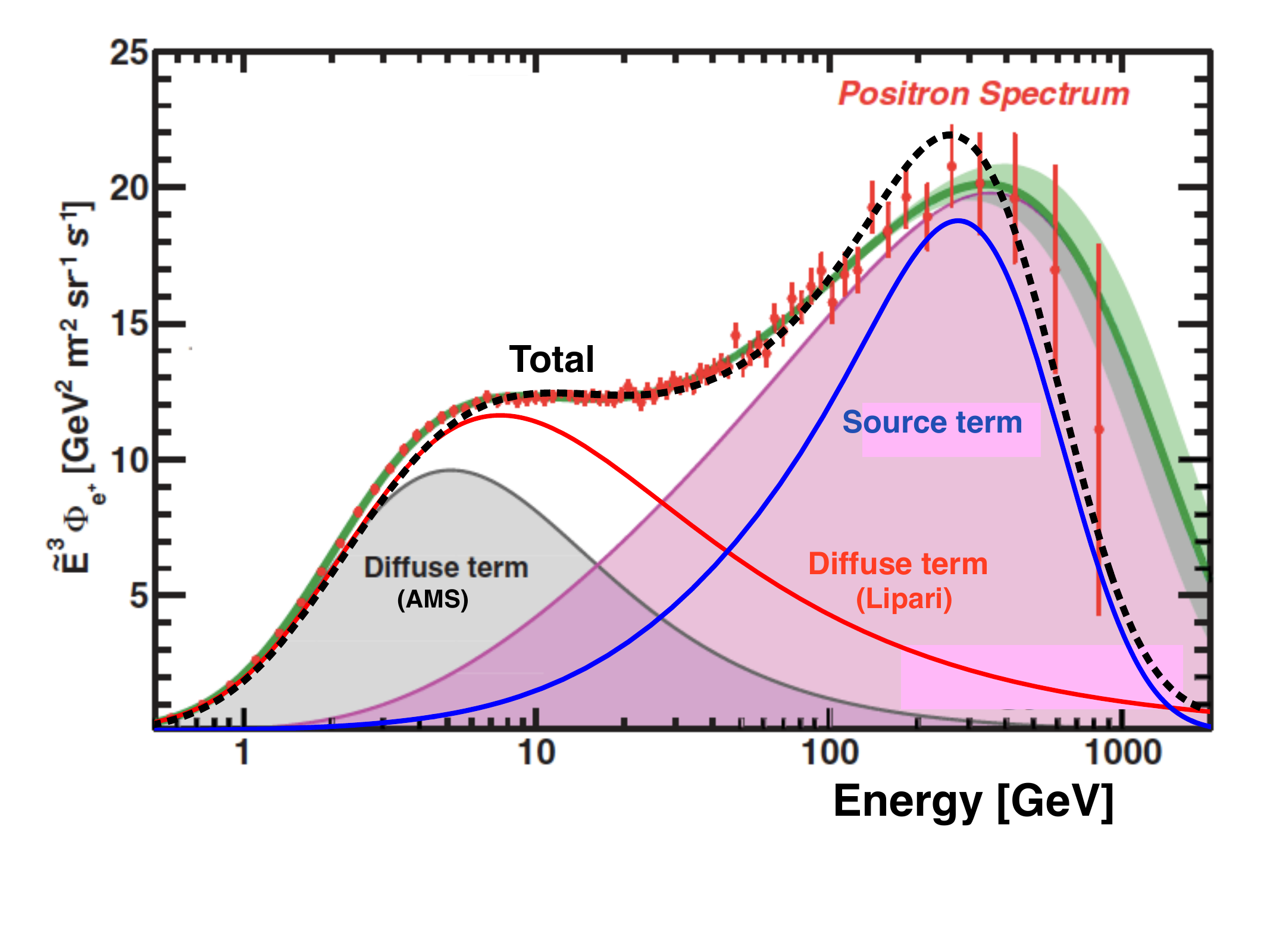}
\vspace{-1.1cm}
\caption{Adopted diffuse term (red) and calculated source term (blue), and their sum
(black, dashed). The calculated source term has been diminished by 20\% \cite{ADReplus}.
The diffuse term used in the sum is the one labelled ``Lipari" \cite{Lipari}.}
\vspace{-.5cm}
\label{fig:Results}
\end{figure}

The result of the complete calculation of the Galactic flux \cite{ADReplus}
of GRB-generated CR positrons is compared
with the AMS data \cite{AMSPos} in Fig.(\ref{fig:Results}). The inputs were 
all typical CB-model parameters or independent items of information, such as the Galactic
rate of GRB-generating supernovas (assumed to be one per century), the two ``diffuse
terms'' (two estimates of the CR $e^+$ secondary CR flux \cite{AMSPos,Lipari}) or the description
of the effects of $e^+$ Galactic confinement-time and energy losses. 

The data preceded the theory, but
the result in Fig.(\ref{fig:Results}) on the shape and normalization of the calculated flux
is either an incredible coincidence or a very satisfactory
consistency check, given the admittedly large number of accumulated uncertainties 
--a non-obvious (i.e., non-linear) example:
an increase by 40\% (or 19\%) of the assumed wind surface density, $\Sigma$, would reduce the
 cited normalization by a factor of two (or 1/1.23).  

The initial angular distribution of the GRB-generated positrons is inmaterial, since their confinement
in the Galaxy's magnetic fields erases all memory of it. For the VHE $\gamma$ rays we shall see
that the angular distribution is crucial and not particularly well known.

\section{Relevant HE laboratory data}

The VHE GRB photons have energies in the range of 100 MeV to a few TeV. They should be produced in individual proton-nucleon collisions at energies at least one order of magnitude higher. As previously
stated, in proton collisions with the light nuclei of the glory the latter approximately
 behave as ensembles of individual protons and neutrons, the two species acting very similarly. 

For our current purposes the only  useful data are collider measurements of photon production
in $pp$ collisions at extremely high peudorapidity intervals; in the relevant center-of-mass system $\eta \equiv -\ln \tan [\theta_{\rm cms}/2]$. Data were gathered at  the LHC at
$\sqrt{s}=13$ TeV \cite{Adriani1}, 900 GeV
\cite{Adriani900} and
510 GeV \cite{Adriani2}. Also relevant are experiments such as \cite{Afanasiev} on Au-Au collisions, that further confirmed the observations of quite precise Feynman scaling: the independence of the results in energy, if 
expressed in terms of the longitudinal and transverse dimensionless 
variables $x\equiv E_\gamma/E_p$ and 
$x_T\equiv 2\,p_T/\sqrt{s}$. 

The collider results from $\sqrt{s}\!=\!0.51$ TeV to 13 TeV 
correspond to a proton beam energy on a proton fixed target of
$\sim\! 270$ TeV and $\sim\! 6700$ TeV, not currently realistic nor directly useful for a CB's protons allegedly generating VHE GRB photons. We would need information for CB's protons on target with energies in the range of $\sim\!1$ to 10 TeV, which we can estimate thanks to the observed scaling laws
and the simplicity of the results.

\subsection{The LHC results}
\label{ss:LHCdata}

\begin{figure}
\centering
\vspace{-.3cm}
\includegraphics[width=9.5cm]{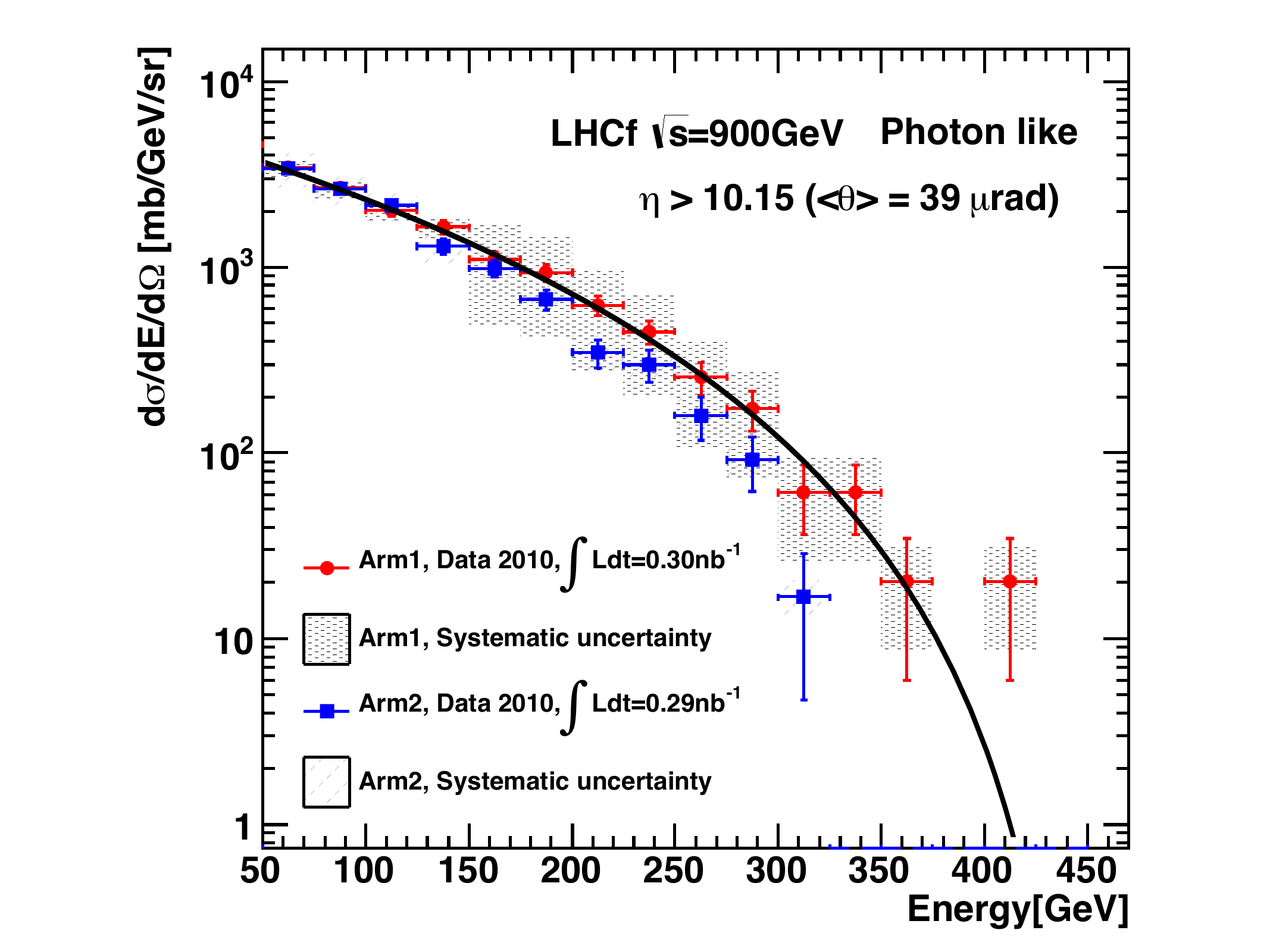}
\caption{The shape of the black line is $\propto\! (1-x)^{3.5}$.
\vspace{-.1cm}
The pseudorapidity is $\eta>10.15$, corresponding to $\theta_{\rm cms} <78$ $\mu$rad.}
\vspace{-.3cm}
\label{Fig:900GeV}
\end{figure}

\begin{figure}[]
\centering
\includegraphics[width=9.5cm]{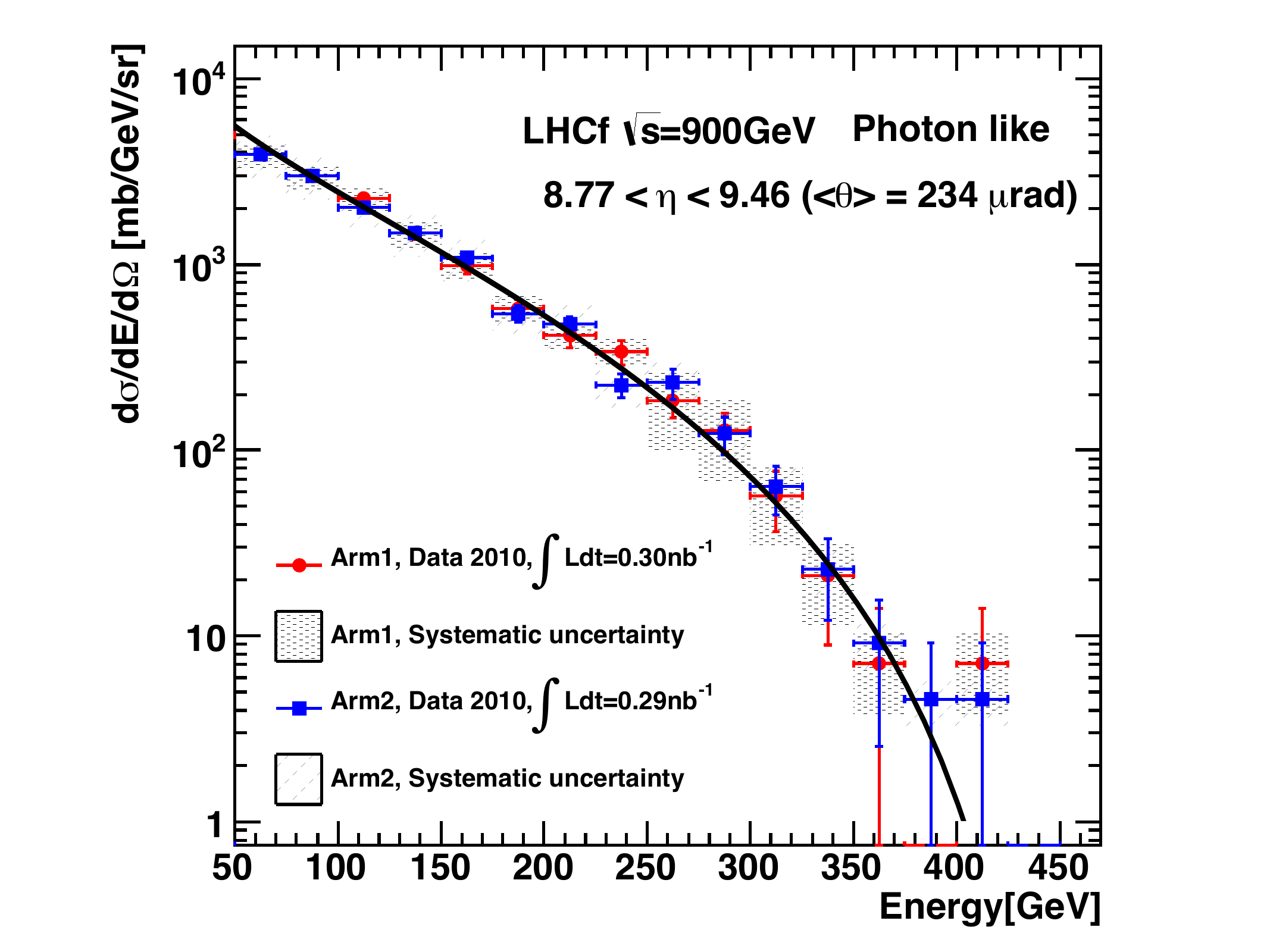}
\vspace{-.6cm}
\caption{The shape of the black line is $\propto\! x^{-0.5}(1-x)^{3.5}$. 
The $\eta$ interval $9.46\!>\!\eta\!>\!8.77$ is $166\!<\!\theta_{\rm cms}\!<\!311$ $\mu$rad.}
\vspace{-.5cm}
\label{Fig:900GeVWide}
\end{figure}

\begin{figure}[]
\centering
\hskip -1.cm
\includegraphics[width=9.5cm]{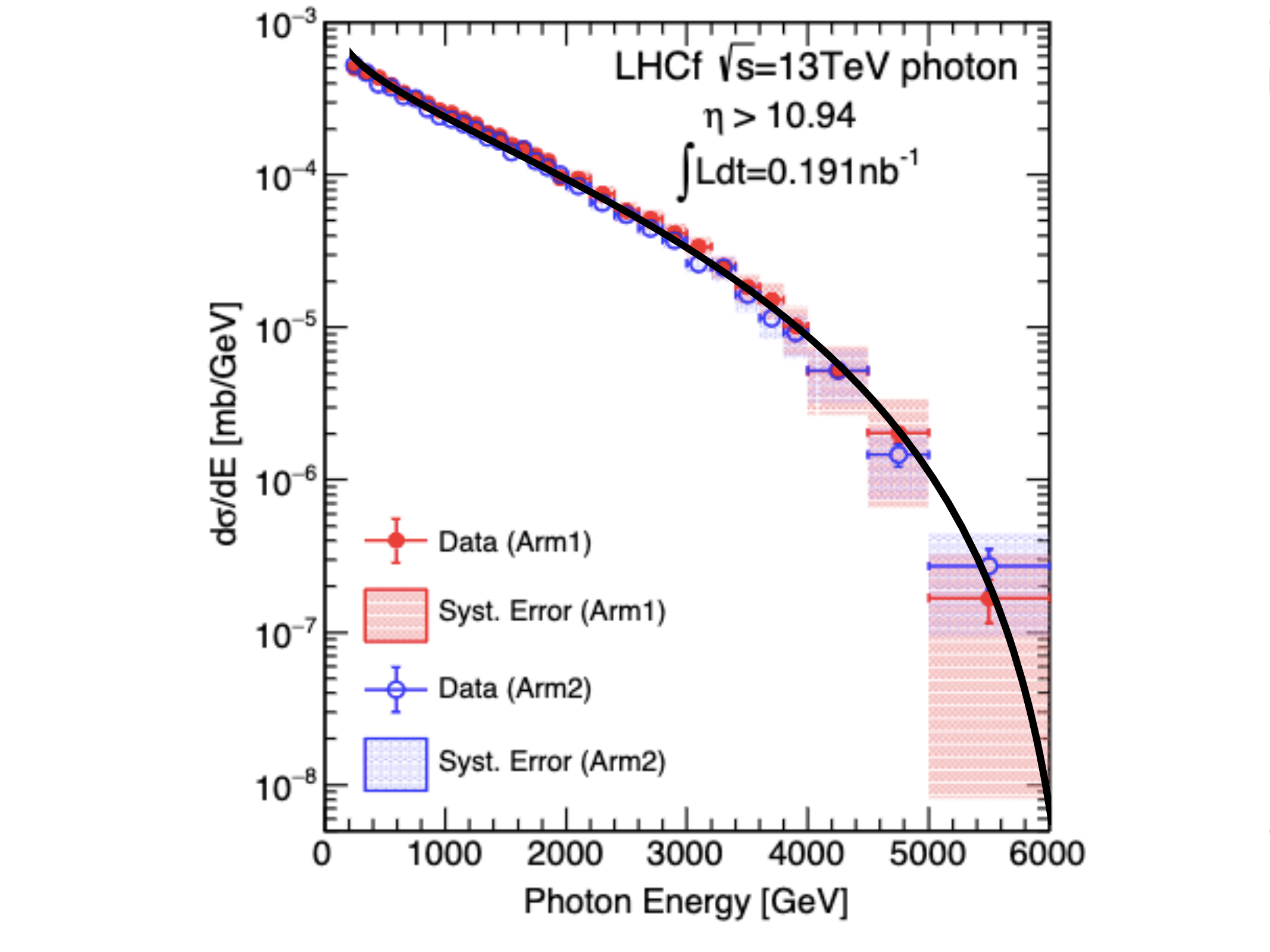}
\vspace{-.5cm}
\caption{The shape of the black line is $\propto x^{-0.3} (1 - x)^{3.5}$. The pseudorapidity 
$\eta>10.15$ corresponds to $\theta_{\rm cms}<78$ $\mu$rad.}
\label{Fig:13TeV}
\end{figure}

\begin{figure}[]
\centering
\hskip -.9cm
\includegraphics[width=9.5cm]{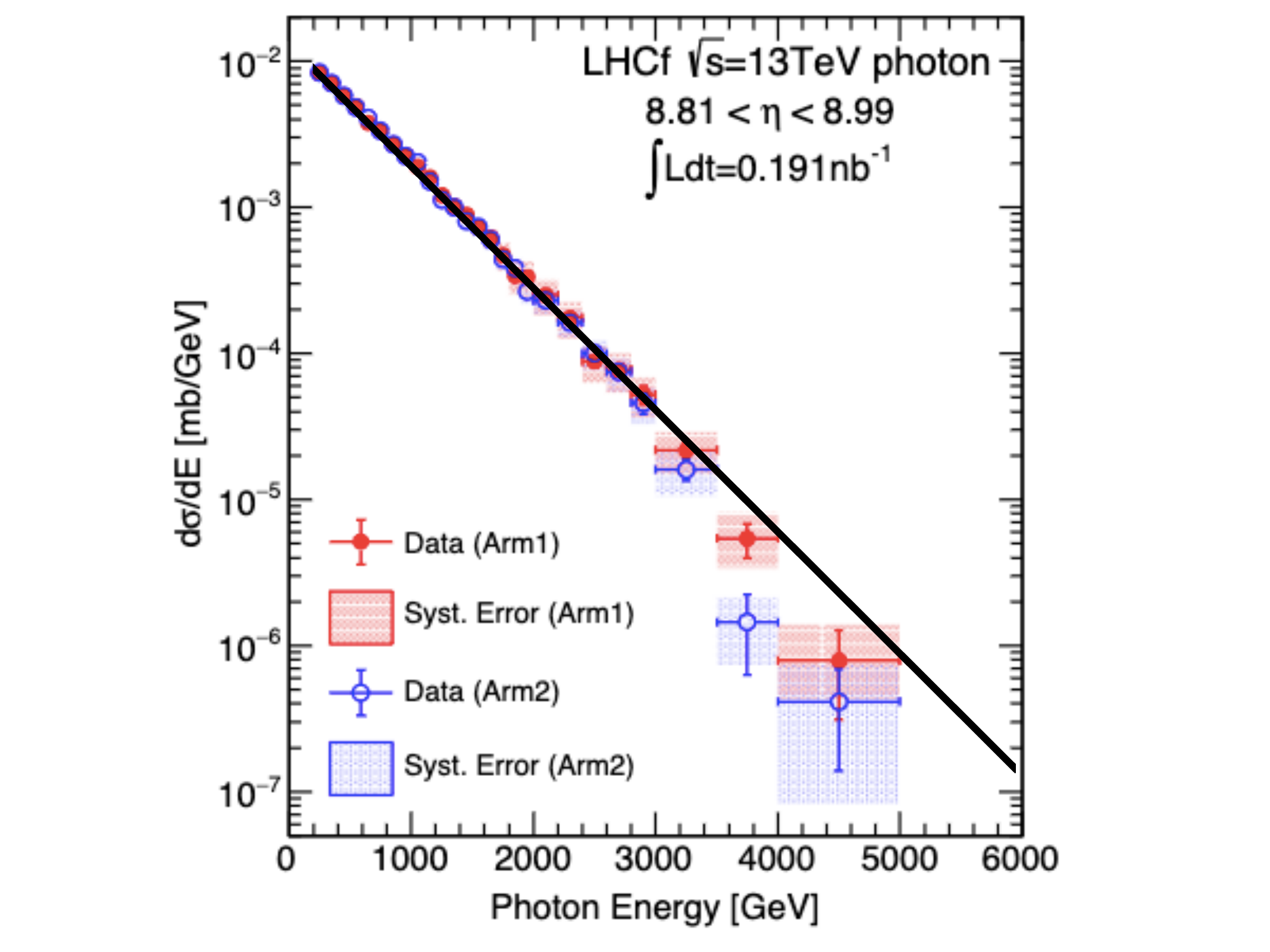}
\vspace{-.5cm}
\caption{The black line's shape is $\propto\!$ Exp$[-12\,x]$. 
The interval $9.46>\eta>8.77$ corresponds to $166<\theta_{\rm cms}<311$ $\mu$rad.} 
\vspace{-.5cm}
\label{Fig:13TeVWide}
\end{figure}

The most relevant measurements are reproduced in Figs.(\ref{Fig:900GeV}) to (\ref{Fig:13TeVWide}).
Fig.(\ref{Fig:900GeV}) is for data at $\sqrt{s}=900$ GeV in a rapidity domain $\eta>10.15$
($\theta_{\rm cms} <78$ $\mu$rad) encompassing the forward direction.
The figure includes a fit of the very simple form,
$(1-x)^{3.5}$, with $x\equiv E_\gamma/E_p$. Data at the same collider's energy is shown 
Fig.(\ref{Fig:900GeVWide}) for the non extremely forward interval 
$9.46\!>\!\eta\!>\!8.77$ ($166\!<\!\theta_{\rm cms}\!<\!311$ $\mu$rad.)
The fit $x^{-0.5}(1-x)^{3.5}$ reflects an enhancement at low $x$: the decays of 
slower $\pi^0$'s and $\eta$'s contribute more significantly at  angles larger
than the ones of Fig.(\ref{Fig:900GeV}).

Results at $\sqrt{s}\!=\!13$ TeV are shown in Figs.(\ref{Fig:13TeV}) and (\ref{Fig:13TeVWide}).
The corresponding rapidity intervals and scaling fits are specified in the figures' captions.
They reproduce the same trends as the $\sqrt{s}\!=\!900$ GeV results. But notice that the 
$\sqrt{s}\!=\!13$ TeV scaling fits differ from the ones at $\sqrt{s}\!=\!900$ GeV. This is not a blatant
breakdown of scaling, it reflects the fact that the rapidity intervals have not been scaled
with energy. It is simpler to change the collider's energy than the locations of forward detectors.

What to conclude from the above discussion in attemting to understand the VHE GRBs
for which we may estimate the energies of the CB's photons but not the angle at which the
CBs, case by case, are viewed? There is one useful tip: approximate the shape of the photon's
energy distribution with a simple one-parameter fit :
\begin{equation}
dE_{\gamma}/dE_p\propto x^\alpha (1-x)^{3.5}.
\label{Eq:x}
\end{equation}

\section{ Analysis of some VHE GRB$\rm s$ }

In a recent ``optimized relativistic fireball model'' study \cite{Foffano} 
the authors do not analyze in detail the $\cal O\rm (250\,MeV)$ 
prompt photons or the afterglow of GRBs, but almost exclusively the VHE $\gamma$ rays
of a few of them. Here we do the same, but in a CB model context.

\begin{figure}[]
\vspace{-0.7cm}
\centering
\includegraphics[width=8.5cm]{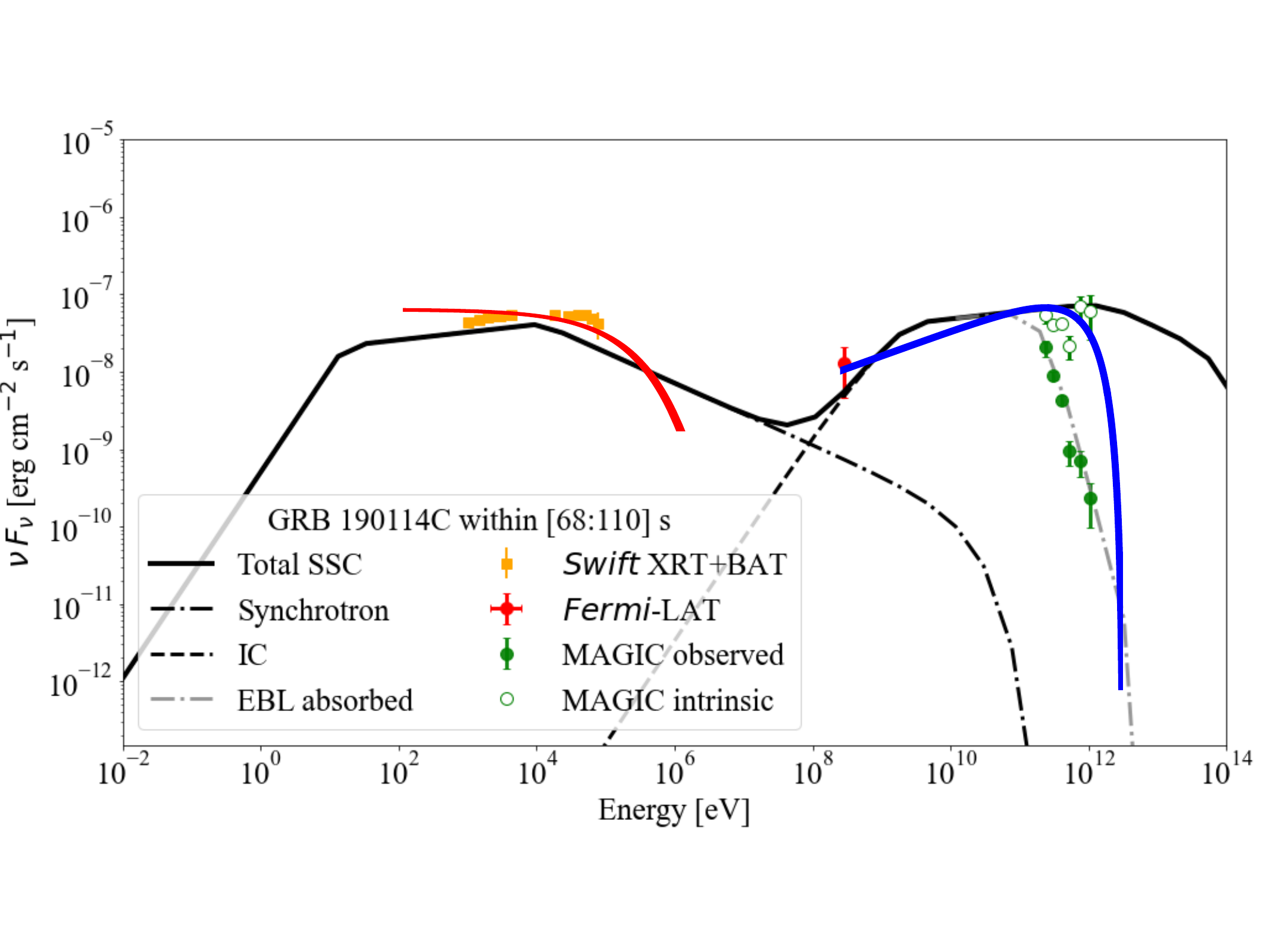}
\vspace{-0.8cm}
    \caption{Figure for GRB190114 from \cite{Foffano}. The red line's shape is 
 Eq.(\ref{Eq:Edistrib}), with $\sigma\!\sim\!4$. The blue line's shape is that of $x^{0.3} (1 - x)^{3.5}$
 with $E_p[CB]\!\sim\!1$ TeV . The hollow points have been un-corrected
 for intergalactic attenuation \cite{Foffano}.}
\vspace{-.6cm}    
\label{Fig:190114C}
\end{figure}

\begin{figure}[]
\vspace{0.3cm}
\centering
\includegraphics[width=8.5cm]{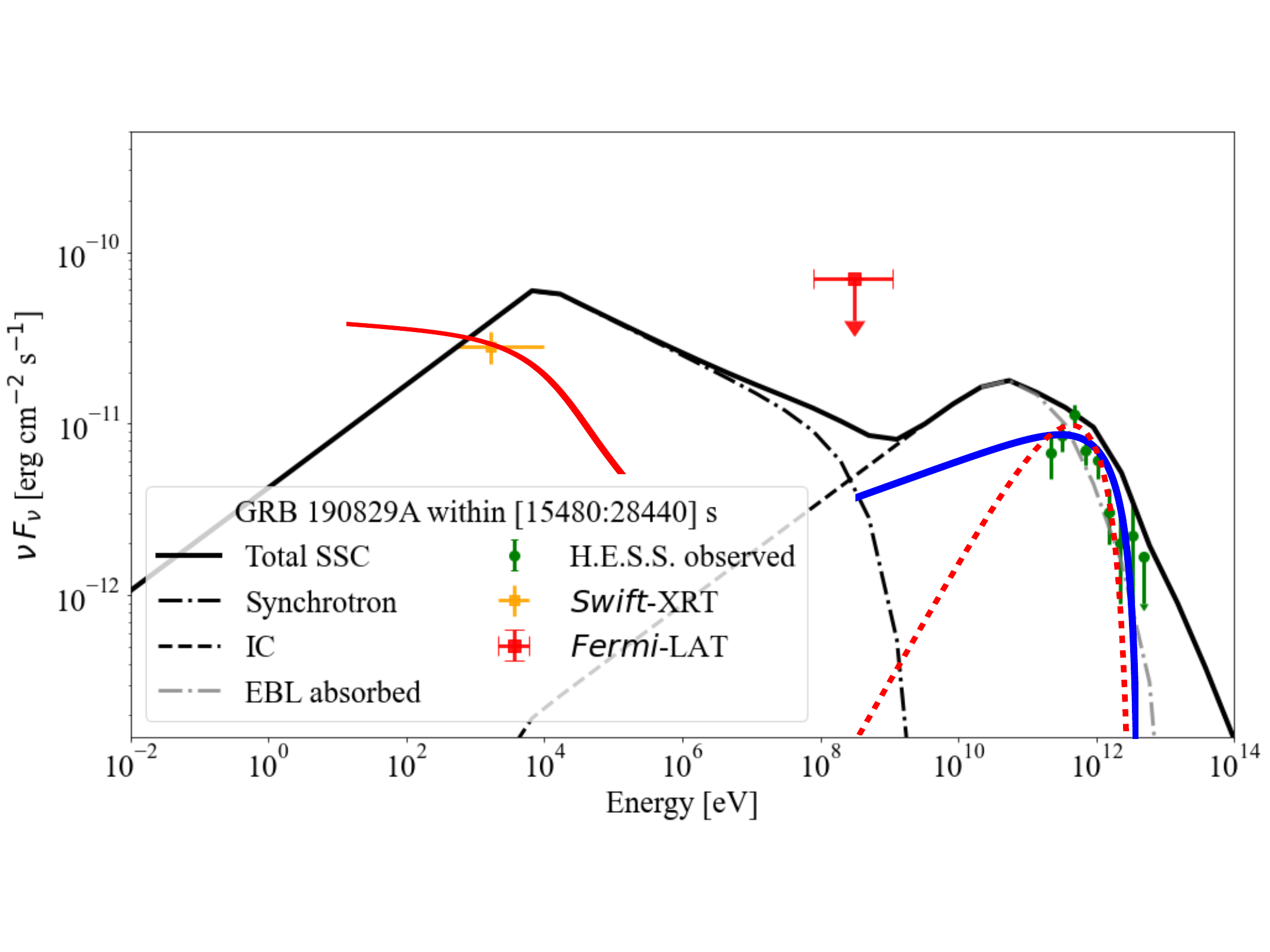}
\vspace{-0.8cm}
\caption{Figure for GRB190829A from \cite{Foffano}. The red line's shape is 
 Eq.(\ref{Eq:Edistrib}), with $\sigma\!\sim\! 1$ but the data around $\sim\!10^3$ eV
are insufficient to constrain the value of $\sigma$.
 The blue line's shape is that of $x^{0.2} (1 - x)^{3.5}$.
 The dotted red line's shape is that of $x(1-x)^{3.5}$.
 The two lines use $E_p[CB]\!\sim\! 3$ TeV.}
 \vspace{-.5cm}   
\label{Fig:190829A}
\end{figure}

\begin{figure}[]
\vspace{-0.3cm}
\centering
\includegraphics[width=8.5cm]{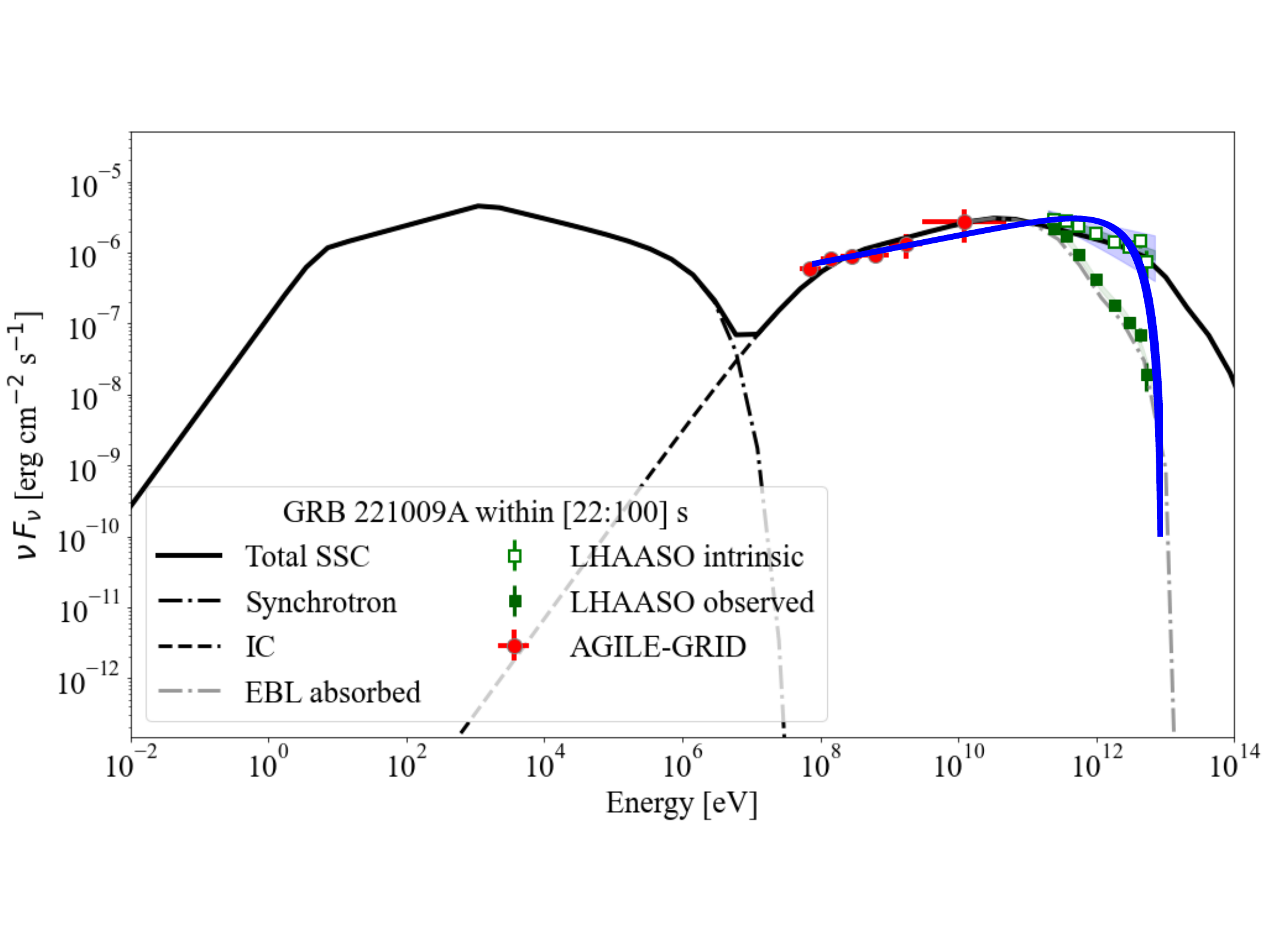}
\vspace{-0.9cm}
\caption{Figure for GRB221009A from \cite{Foffano}. 
The blue line's shape is that of $x^{0.2} (1 - x)^{3.5}$.
 with $E_p[CB]\!\sim\! 10$ TeV. 
 The hollow dots were un-corrected for photon attenuation \cite{Foffano}.}
 \vspace{-1.5cm}  
\label{Fig:221009A}
\end{figure}

In Figs.(\ref{Fig:190114C} to \ref{Fig:221009A}), 
we show results for some
representative VHE GRBs, namely 190114C, 190829A and 221009A. The figures are taken from
\cite{Foffano}. Added, in red, 
to Figs.(\ref{Fig:190114C}) and (\ref{Fig:190829A})
are rough expectations from Eq.(\ref{Eq:Edistrib}).
Also added to these figures (in blue) are the VHE CB-model description of the data with
$x$-distributions as in Eq.(\ref{Eq:x}) with $\alpha=$ 0.3, 0.3 and 0.2, 
respectively. The approximate energies $E_p(CB)$ of the protons in the CBs that produced
the photons are are 1, 3 and 10 TeV.
The dashed red line in Fig.(\ref{Fig:190829A}) is the result for $\alpha=0$,
showing the extreme sensitivity to the value of $\alpha$.
According to our discussion of the lowest-energy LHC data of Fig.(\ref{Fig:900GeV}) 
$\alpha=0$ would reflect a very small photon viewing angle.

\section{Timing considerations}

Let the measurement of a property of a GRB last for a time $dt_{\rm Obs}$.
In this time a CB allegedly traveling at $v\!\simeq\! c$ and emitting the observed radiation has, in the rest system
of the host galaxy, travelled a distance $dx_{\rm HG}$:
\begin{equation}
dx_{\rm HG}={{\gamma\,\delta} \over {1+z}}\,c\,dt_{\rm Obs}
\label{Eq:distance}
\end{equation}
For a typical good old fashioned GRB, with $\sigma\!=\!1$ in Eq.(\ref{Eq:boosting}), 
$dt_{\rm Obs}\!=\! 1$ s means $dx_{\rm HG}\!=\!10^6$ s. The CB has travelled for
almost twelve days, squeezing into one second the information it emits of the properties of the
medium it traverses. This is even more dramatic a fast-forward video for VHE 
($\sigma\!\gg\!1$) GRBs.
Wonders of relativity.

In the CB model two mechanisms produce significant ``structure'' in the time sequence of 
the observations. One, already mentioned, 
 is the evidence for more than one CB, producing the matching
number of ICS peaks in the ``prompt'' photon-counting rate. 

\begin{figure}[]
\centering
\hskip -.35cm
\vspace{.2cm}
\includegraphics[width=9cm]{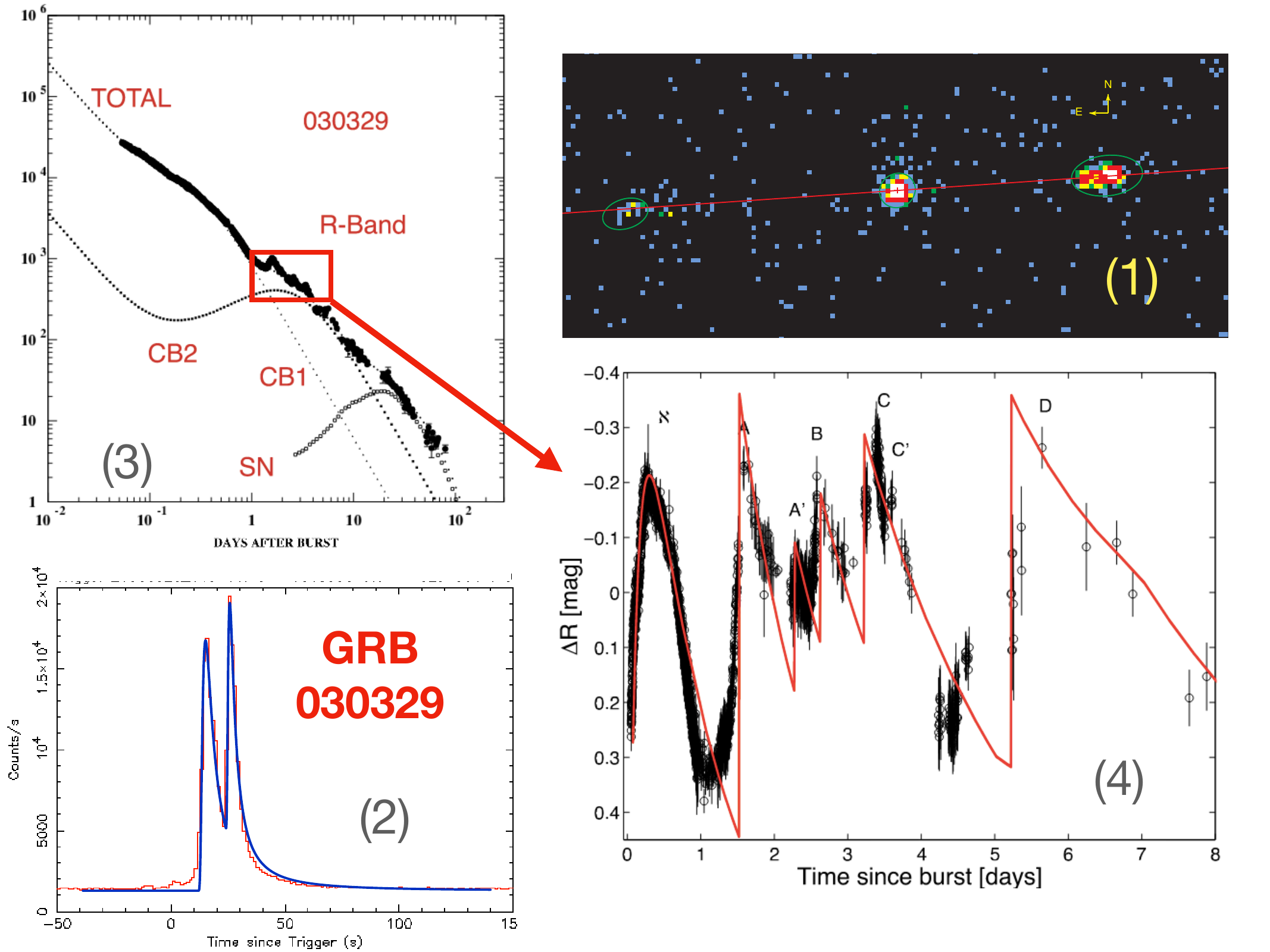}
\caption{(1)  XTE J-1550-564 as two opposite simultaneously 
emitted CBs re-brighten. (2) The two-CB prompt pulse of GRB030319. 
(3)  Its R-band afterglow.
(4) A blow-up of the
observed structures on days 1 to 6 after burst. }
\vspace{-.5cm}
\label{Fig:overdensities}
\end{figure}

A second source of structure is due
to variations in the ISM density through which a CB travels. These manifest 
themselves as rebrightenings in the GRB's AG, entirely
analogous to the ones observed and well understood in the case of the microquasar
XTE J-1550-564 \cite{microquasar}. And then, in the late AG, naturally, there is the light from the
SN that ejected the CBs\footnote{The  enthralling ``science'' and sociology of the GRB/SN association
is narrated in \cite{ADRupdate} and, more colloquially, in \cite{Coll}.}.
GRB030329, 
had two clear peaks \cite{030329},  shown in Fig.(\ref{Fig:overdensities}).
In the AG their contributions are labelled CB1 and CB2 in the figure.
We predicted  the night in which the SN was to be observed  \cite{030329} but our prediction for the
superluminal separation between the CBs was wrong until we modeled 
the effects \cite{030329a}
of the various over-densities as in Fig.(\ref{Fig:overdensities}).

Data on five GRBs are gathered in \cite{Foffano}. GRB180720 displays pronounced peaks up to 200 s after ``$T_0$''. They reflect ICS by a multitude of CBs and/or the presence of several ICS-enhancing pre-SN ejections. Similar though less pronounced structures are seen in 
GRB190114C up to $\sim\! 6$ s after $T_0$.
The VHE $\gamma$-rays of GRB190829A were seen 4.3 to 55.9 hours after $T_0$.
These must be due to the effect of ISM density variations on the synchrotron-radiation AG.
The same can be said about the VHE $\gamma$s of GRB 201216C and 221009C, which were 
respectively seen 20 minutes and 28 hours and after $T_0$.

\section{No HE neutrinos}
\label{Sec:Nonus}

VHE $\gamma$'s and neutrinos are expected to be co-produced by similar mechanisms
\cite{OldNeutrino-expectations}, why have the GRB-associated neutrinos not been seen?

At high energies, due to isospin symmetry, the $pp$ and $pA$ collisions produce $\pi^+$s,
$\pi^-$s and $\pi^0$s in similar amounts and transverse-momentum distributions. 
Muons in $\pi\to \mu\nu$ decays carry most of the pion energy and momentum. The three-body decays
$\mu\to e\nu\bar\nu$ result in neutrinos
 with transverse momenta distributions $\sim \! 2/3$ softer than the ones of the 
$\gamma$'s of the two-particle decay $\pi^0\!\to\! \gamma\gamma$ decay. At 
small angles and the highest energies the production of neutrinos is dominated by 
charmed-particle production and decay, a 
verified \cite{CharmedNeutrinos} prediction \cite{CharmedNeutrinos2}.
Due to neutrino oscillations the original neutrino beams become an equal
admixture of $\nu_e$, $\nu_\mu$ and $\nu_\tau$. 

It is possible to elaborate in detail all items in the previous paragraph. But the useful
conclusion is that the beam of HE GRB neutrinos has a similar energy distribution and opening
angle as that of HE gamma rays.

 Are the GRB neutrinos detectable? The
 $\nu_\mu$ cross section on water at 1$\!$ TeV is $1.35 \!\times\! 10^{-34}$
cm$^2$/molecule. The photon cross section at the same energy in the high-$Z$
materials used in some satellite detectors reaches up to
$\sim\! 10^{-24}$ cm$^2$/molecule. It is this ten orders of magnitude difference that
the large-volume water Cherenkov detectors must compensate for and, apparently, 
they have not.

\section{Conclusions}

The CB-model understanding of the HE $\gamma$ rays of some GRBs
is not nearly as precise as the understanding in the same model of the 
${\cal O}(250\,\rm keV)$ prompt radiation and the GRB afterglows. In the near
future it will not be.  We lack
precise direct information on the angle $\theta$ 
at which GRB HE $\gamma$ rays are observed and on their Lorentz factors, though
this is in principle extractable from the prompt radiation and afterglow of each
GRB. To boot, we lack the required
information on the proton-nuclei production of HE $\gamma$ rays at the relevant
energies and  angles (compare this to the understanding of ICS required
to derive all properties of the prompt radiation!).

Yet, the CB-model, as argued, has no difficulty in describing the origin and
approximate properties of the HE $\gamma$ rays and the extreme difficulty
of observing their accompanying neutrinos.

\vspace{.2cm}
{\bf A multi-messenger afterthought}
 
 Should the next supernova in our side of the Galaxy be of the proper Type, there is no question
 that its CBs would be easily observable. It is even possible that the successive episodes
 of accretion --that allegedly trigger CBs-- involve high-enough densities for neutrinos to
 be emitted, as discussed in \cite{1987Anus}, a proposal that did not receive a warm
 approval. Successive neutrino emissions would manifest themselves as subdominant
 peaks in the neutrino flux as a function of time.


{\bf Acknowledgements.} 
I am indebted to Shlomo Dado and Fabio Truc for advice and encouragement.
This project has received funding from the European Union's Horizon 2020 research and innovation programme under the Marie Sklodowska-Curie grant agreement No 674896.


\end{document}